\documentstyle[11pt,newpasp,twoside,epsf]{article}
\def\edcomment#1{\iffalse\marginpar{\raggedright\sl#1\/}\else\relax\fi}
\reversemarginpar

\begin{document}
\title{Hard X-ray Observations of Magnetic Cataclysmic Variables}
 \author{K. P. Singh, V. R. Rana and K. Mukerjee}
\affil{Tata Institute of Fundamental Research, Homi Bhabha Road, Colaba, 
Mumbai 400005, INDIA }

\author{P. Barrett}
\affil{Space Telescope Science Institute, ESS/Science Software Group, Baltimore, MD 21218, U.S.A.}
\author{E. M. Schlegel}
\affil{Harvard-Smithsonian Center for Astrophysics, 60 Garden Street, Cambridge, MA 02138, U.S.A.}

\begin{abstract}
Hard X-ray light curves and spectral parameters from our analysis of 
X-ray data of five AM Her type systems -- V2301 Oph, V1432 Aql, EP Draconis,  
GG Leonis, \& V834 Cen, and one intermediate polar -- TV Col, observed using
the $Rossi~X-ray~Timing~Explorer$ $(RXTE)$ satellite are presented.
A new improved ephemeris has been derived for V2301 Oph using the mid-eclipse
timings.  Average intensity variations, without any change of shape of the
light curve or hardness ratio, are observed on timescales of a few days to a 
few months in V2301 Oph.  V1432 Aql shows erratic variations on a timescale
of a day, at least 2 sharp dips near orbital phases 0.35 and 0.5, and a 
total eclipse.  Hard X-ray eclipses are also reported in EP Dra and GG Leo.  
V834 Cen shows intensity variations on yearly timescale and is found
to be in a low state in 2002.
In TV Col, a binary orbital modulation at 5.5h, in addition to the 
spin period of 1910s, is reported for the first time. 
Maximum spectral temperatures in Polars have been determined 
and used to estimate the masses of the white dwarfs.  

\end{abstract}

\section{Introduction}
Hard X-ray ($>$ 2 keV) emission from polars is believed to originate from 
the post-shock region at the base of the accretion column 
(see Patterson 1994) on the accreting white dwarf.  
X-ray studies of eclipsing polars are particularly valuable for
understanding the geometry of the accretion region, and to learn about the 
interaction of matter with the magnetic field near
the coupling region, the stellar masses, the shock height, and the
temperature distribution near the white dwarf surface.  
A hard X-ray study of polars with large area and broad-band X-ray detectors 
like the Proportional Counter Array (PCA) on board the ($RXTE$) satellite 
provides an excellent opportunity to learn about these effects and to 
estimate the mass of the white dwarf by measuring the maximum temperature 
of the post-shock plasma.

We have analyzed the X-ray data obtained of 5 Polars and 1 Intermediate Polar 
(IP) as listed in Table 1 along with some of their important properties.  
The dates of the observations of each 
source and the mean count rate in the 2--20 keV band are given in Table 2.   
Most of the observations have the source in the center of the field of 
view of PCA, except for the 2002 observation of V1432 Aql which is 
intentionally offset to minimize the contribution of flux from NGC6814.
\begin{table}
\caption{Properties of the 5 Polars \& 1 IP in the sample}
\begin{tabular}{llllll}
\tableline
Name  & Magnitude & Spin Period & Distance &  References\\
 & (Visual) & 	(seconds)	& (pc) & \\
\tableline
V2301 Oph & 15--17 & 6780 & 150$\pm$27 & SRHB, BRB, Hess97, SI, SS\\
V1432 Aql & 14.2--18 & 12150 & 230 & W95, P95, GS\\
EP Dra & 18 & 6276  & $\sim$450 & RSTS, S99, SM\\
GG Leo & 16--17 & 4792.767 & $>$100 &  Bur98, Sz00\\
V834 Cen & 15 & 6090 & 86 & JNJ, M83, B83, C90, S93\\
TV Col & 14.1 & 1911 & 368 & BDW, H93, A94, R02\\
\tableline
\tableline
\end{tabular}
\end{table}
\begin{table}
\caption{Summary of $RXTE$ Observations analyzed}
\begin{tabular}{lllll}
\tableline
Name & 	Date &  Exposure Time &   Count Rate &  Offset(\arcmin)\\
 & (Y M D) &   (seconds)      &  (2-20 keV) & \\
\tableline
V2301 Oph & 1997 05 27-30 & 13200 & 6--16 & 0.06\\
          & 1997 09 26-28 & 5700 & 4-21 & 0.06\\
          & 1997 11 22-24 & 39500 & 15-23 & 0.06\\
V1432 Aql & 1998 06 23-25 & 16700 & 23-33 & 0.01\\
          & 2002 07 14-15 & 19700 & 18-20 & 14.8\\
EP Dra    & 1998 08 28    & 14303 & 1.2 & 0.1\\
GG Leo    & 1999 05 07    & 4688  & 3.3 & 0.03\\
V834 Cen  & 1996 05 14-18 & 8700  & 5-9 & 0.01\\
          & 1997 08 02    & 16256 & 8 & 0.03\\
          & 1998 06 23-25 & 20400 & 9-11 & 0.04\\
          & 2002 01 18, 24 & 18600 & $<$ 0.5 & 0.04\\
TV Col    & 1996 08 09-13 & 82000 & 30-50 & 0.1\\
\tableline
\tableline
\end{tabular}
\end{table}
\begin{figure}
\plotone{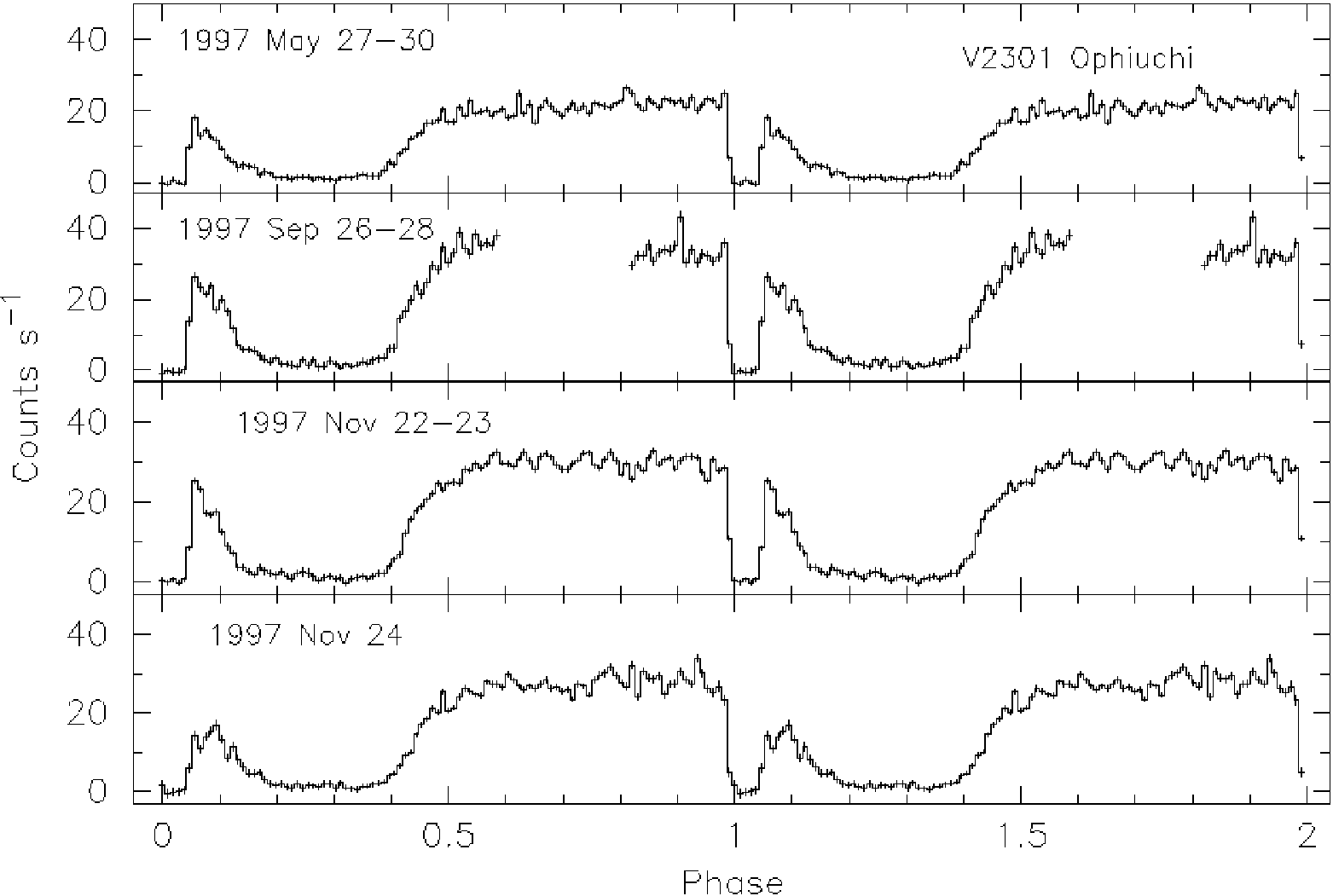}
\caption{X-ray lightcurve of V2301~Oph shown folded on our improved ephemeris.  The bin size is 64~s.}
\plotone{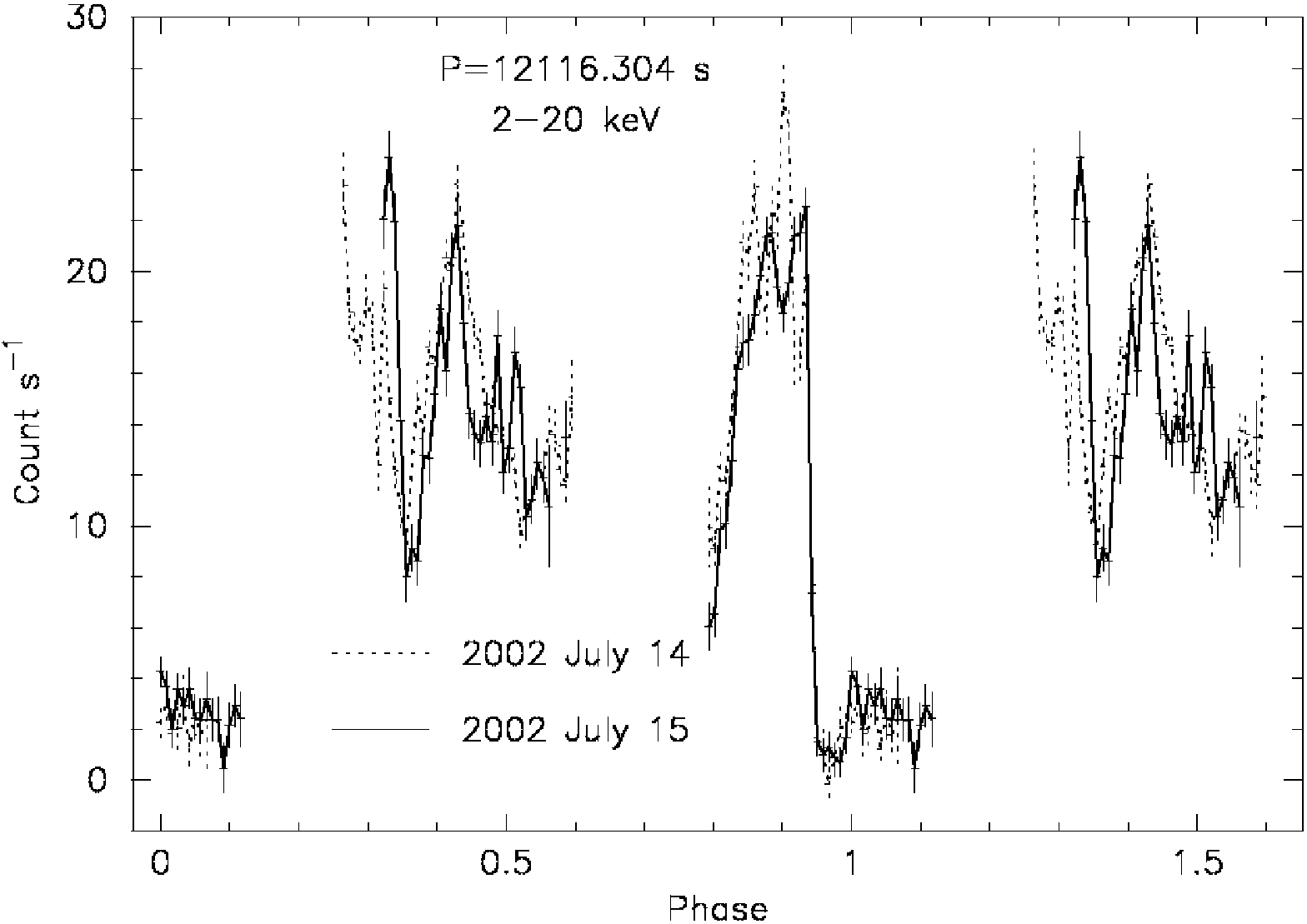}
\caption{X-ray lightcurve of V1432~Aql shown folded on its orbital period (Ephemeris taken from W95). The bin size is 64~s.}
\end{figure}
\begin{figure}
\plotone{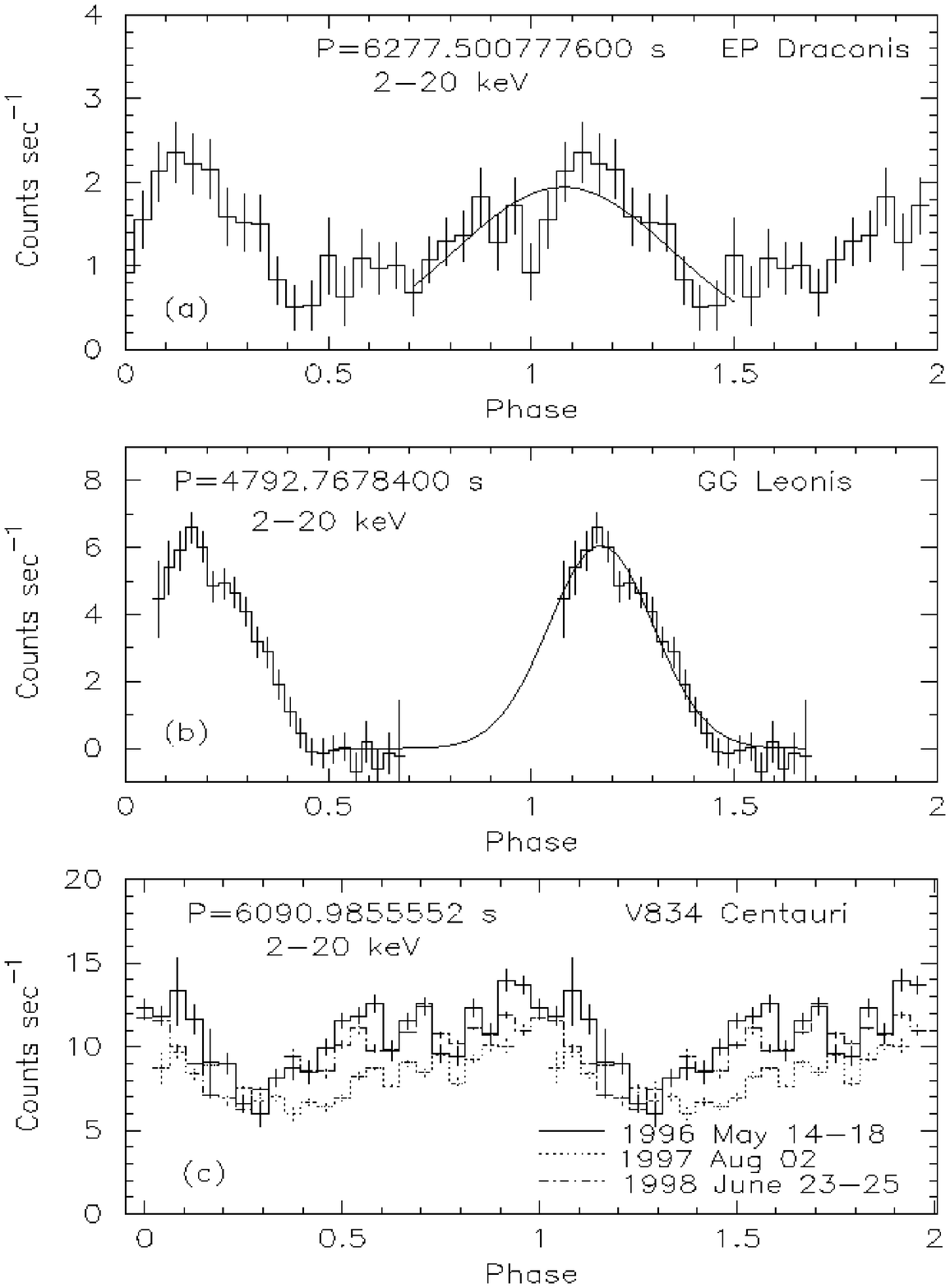}
\caption{X-ray lightcurves of EP~Dra, GG~Leo \& V834~Cen shown folded on their orbital periods (Ephemeris are from SM, Sz00 \& S93 respectively).}
\end{figure}
\section{2-20 keV Light Curves of Polars}
X-ray light curves of V2301~Oph and V1432~Aql in the 2-20 keV 
energy band are shown in Figures 1 and 2 respectively 
for each of the long observations.  Several total eclipses 
were seen clearly in V2301 Oph and along with previous
observations used to derive an improved 
ephemeris for the system: HJD = 2448071.02051(2) + 0.0784499794(10)E.
The out-of-eclipse intensity of V~2301 Oph changes on daily and monthly 
timescales, while the light curve shape and the eclipse ingress and
egress remain unchanged (see Fig. 1).  
The spectral hardness ratio is, however, found to remain constant throughout. 
These intensity variations, therefore, lead to luminosity variations and
suggest that the mass accretion rate varies on these timescales.  
V1432 Aql also shows complete eclipses 
(the small residual amount $\leq$10\% of the peak is due to
contamination by NGC~6814) in both the observations shown in Fig. 2. 
Several dips are seen between the phases 0.35 to 0.52, and 
rapid flickering is observed near the maximum in the highly 
complicated light curve of V1432~Aql.  Total eclipses have also been observed
in the 2--20 keV light curves of EP Dra and GG Leo shown in Fig. 3.  
The residual
count rate in EP~Dra is fully accounted for by contribution from a nearby 
($\sim$8.3\arcmin away) cluster of galaxies A2317.  
V834~Cen shows yearly intensity variations (Fig. 3) with changing maximum 
intensity as a function of phase while the minimum intensity 
remains the same.  
It was not detectable in 2002, and therefore, presumed to have 
a entered a very low state.
\section{TV Col: Spin and Orbital Periods}
A 5 day long X-ray observation of an intermediate polar, TV Col, 
is reported here for the first time.  
The power spectral distribution of its light curve obtained
by using the CLEAN (see Norton, Beardmore \& Taylor 1996)
method and thus removing the window function,
is shown in Fig. 4.  It clearly displays the presence of a sinusoidal 
oscillation (spin period) at 1910s and that of modulation due to 
orbital period of 5.5h. 
Several other peaks due to various harmonics and side bands of the spin,
orbital period and the precession period of 4d (see Norton et al. 1996; 
Retter et al. 2002) are also seen.  In particular, power is detected
at side-band frequencies corresponding to periods of 5.2h, 5.8h, and 6.2h 
(or beats between the orbital and precession period).
\begin{figure}
\plotfiddle{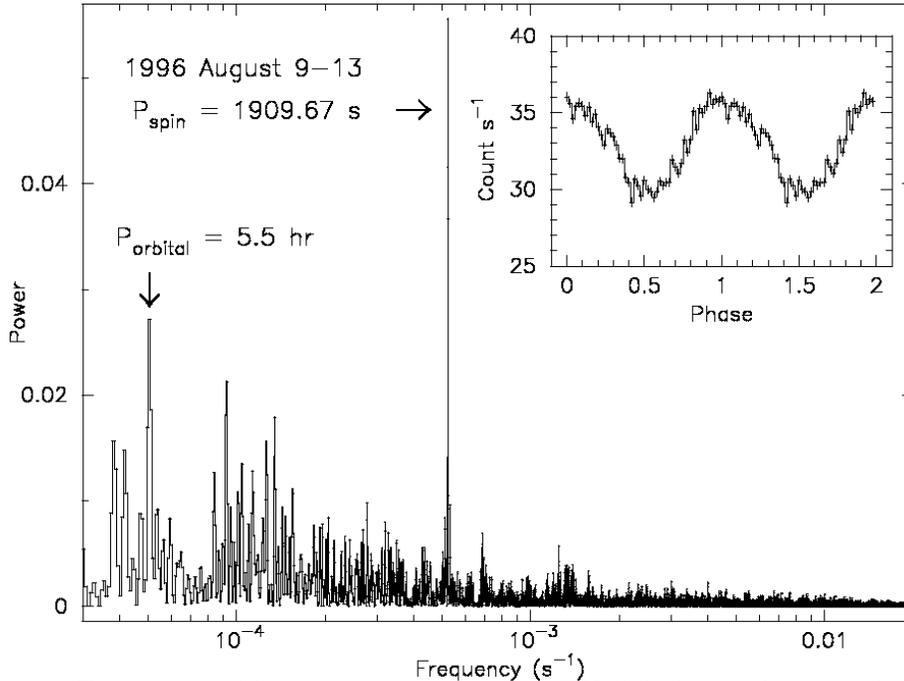}{3.0in}{0}{53}{53}{-200}{-22}
\caption{Clean power spectrum of TV~Col obtained after subtracting the 
mean count rate. Inset shows modulation at the spin period.}
\end{figure}
\section{X-ray Spectra of Polars}
The hard X-ray emission is believed to originate from the hot gas in 
the post-shock region gradually cooling and settling on to the white 
dwarf.
It is, therefore, expected that the post-shock region should have a
temperature gradient with a low temperature at the bottom of the accretion
column and high near the shock front, and a continuous temperature 
distribution in between.  Multi-temperature plasma models are therefore 
required to quantify the temperature distribution and line emission 
(Done, Osborne \& Beardmore 1995).  In addition, the pre-shock
material is exposed to hard X-rays thus ionizing it, which can give rise to
complex absorption effects.   A considerable fraction of hard X-rays from
the post-shock region, can also be reflected by the white dwarf surface.
Hence, it is important to take into account all the above processes
while characterizing the X-ray emission.  Several spectral models 
ranging from the simple thermal bremsstrahlung to plasma emission 
models with continuous emission measure (cevmekl), and cooling flow 
were used to fit the X-ray spectra.  
The best fit spectral parameters derived from two models : thermal 
bremsstrahlung plus absorption, and cevmekl with ionised absorber 
and reflection are shown in Table 3.
The 2--10 keV flux values given in Table 3 are in units of 
10$^{-11}$ ergs cm$^{-2}$ s$^{-1}$.  The errors and the range of values
for the parameters listed in the table are with 90\% confidence. 
The thermal bremsstrahlung models always required the presence 
of an emission line component (a broad Gaussian) with the line energy 
indicating the dominance of the plasma
emission from Fe XV or FeXVI over fluorescence of cold Fe. The cevmekl models
describe the physical situation much better. These models indicate subsolar 
abundance for Fe and provide a maximum temperature that is somewhat lower 
than that derived from the simple bremsstrahlung models (Table 3). The
best fit maximum temperatures from Model 2 were used to estimate 
the mass of the white dwarf in these systems.  
The estimated masses (Table 3) are consistent, 
within errors, with the previous estimates given by Ramsay (1997) 
using a more realistic spectral model. 
\begin{table}
\caption{Spectral Parameters of Polars}
\begin{tabular}{lllllllll}
\tableline
      & Model 1 & & & & Model 2 & & \\
Name & kT$_{brem}$ & E$_{Fe}$ & kT$_{max}$ &  N$_{H}$(warm) & A$_{Fe}$ & Flux$^a$ & M$_{WD}$ \\
     &  keV & keV &  keV & 10$^{22}$cm$^{-2}$ & Relative &  & M$_{\sun}$\\
\tableline
V2301 Oph & 12-15 & 6.5-6.8 & 10-20 & 1.5-3.2 & 0.11-0.28 & 3.30 & 0.90$^{+0.10}_{-0.15}$\\
V1432 Aql & 50-75 & 6.4-6.6 &  12-18 & 2.4-3.5 & 0.17-0.31 & 4.00 & 0.90$^{+0.08}_{-0.07}$\\
EP Dra & 5-9 & 6.5-7.1  & 10$^{+5}_{-4}$ & $<$1.1 & 0.3-1.4 & 0.32 & 0.75$^{+0.18}_{-0.32}$\\
GG Leo & 38$^{+56}_{-16}$ & 6.4-7.0 & 22$^{+78}_{-15}$ & $<$5.0 & 0.15-0.45 & 0.84 & 1.1$^{+0.3}_{-0.5}$\\
V834 Cen & 11$^{+2}_{-2}$ & 6.6-7.0 & 8.5$^{+4.5}_{-3.5}$ & 1.5-3.8 & 0.20-0.67 & 1.66 & 0.70$^{+0.15}_{-0.20}$\\
\tableline
\tableline
\end{tabular}
\end{table}


\begin{references}
Augusteijn, T., Heemskerk, M.H.M., Zwarthoed, G.A.A., \& van Paradijs, J. 1994, \aaps, 107, 219 (A94)

Bailey, J., Axon, D. J., Hough, J. H., Watts, D. J., Giles, A. B., \& Greenhill J. G., 1983, \mnras, 205, 1P (B83)

Barrett, P., O'Donoghue, D., \& Warner, B. 1988, \mnras, 233, 759 (BDW)

Barwig, H., Ritter, H., \& B$\ddot{a}$rnbantner, O., 1994, \aap, 288, 204 (BRB)


Burwitz, V., Reinsch, K., Schwope, A. D., et al., 1998, \aap, 331, 262 (Bur98)

Cropper, M., 1990, \ssr, 54, 195 (C90)


Done, C., Osborne, J. P., \& Beardmore, A. P., 1995, \mnras, 276, 483.







Geckeler, R. D. \& Staubert, R., 1997, \aa, 325, 1070 (GS)

Hellier, Coel, 1993, \mnras, 264, 132 (H93)

Hessman, F. V., Beuermann, K., Burwitz, V., de Martino, D., \& Thomas, H. C., 1997, \aap, 327, 245 (Hess97)


Jensen, K. A., Nousek, J. A., \& Nugent, J. J., 1982, \apj, 261, 625 (JNJ)


Mason, K., O., Middleditch, J., Cordova, F., A., et al., 1983, \apj, 264, 575 (M83)

Norton, A. J., Beardmore, A. P. \& Taylor, P., 1996, \mnras, 280, 937

Patterson, J., 1994, \pasp, 107, 307

Patterson, J., Skillman, D. R., Thorstensen, J. \& Hellier, C., 1995, \pasp, 107, 307 (P95)

Ramsay, G., 1997, \mnras, 290, 99

Remillard, R. A., Stroozas, B. A., Tapia, S. \& Silber, A., 1991, \apj, 379, 715 (RSTS)

Retter, A., Hellier, C., Augusteijn, T., Naylor, T., et al., 2002, \mnras, in press (R02)

 
Schlegel, E. M., 1999, \aj, 117, 2494 (S99)

Schmidt, G. D., \& Stockman, H. S., 2001, \apj, 548, 410 (SS)

Schwope, A., D., \& Mengel, S., 1997, Astron. Nachr., 318, 25 (SM).

Schwope, A., D., Thomas, H., C., Beuermann, K., \& Reinsch, K., 1993, \aap, 267, 103 (S93)

Silber, A. D., Remillard, R. A., Horne, K. A. \& Bradt, H. V. 1994,\apj, 424, 955(SRHB)

Steiman-Cameron, T. Y., \& Imamura, J. N., 1999, \apj, 515, 404 (SI)

Szkody, P., Armstrong, J., \& Fried, R., 2000, \pasp, 112, 228 (Sz00).

Watson, M. G., Rosen, S. R., O'Donoghue, D., et al., 1995, \mnras, 273, 681. (W95)
\end{references}
\end{document}